\documentclass[CEJP,DVI]{cej}
\usepackage[utf8x]{inputenc}
\usepackage[T1]{fontenc}
\usepackage{graphicx}
\usepackage[labelfont=bf,font=small,format=hang,figureposition=below,tableposition=above,labelsep=quad]{caption}
\usepackage{layout}
\usepackage{amsmath,amsfonts,amssymb}
\usepackage{textcomp}
\usepackage{hyperref}
\usepackage{mathtools}
\usepackage{calc}

\title{Wave Propagation And Landau-Type Damping In Liquids}

\articletype{Research Article}

\author{Vincenzo Molinari\inst{1},
        Domiziano Mostacci\inst{1}\email{domiziano.mostacci@unibo.it}}

\institute{
     \inst{1} Laboratorio di Montecuccolino, Alma Mater Studiorum Università di Bologna,\\
     Via dei Colli 16, I-40136 Bologna, Italy
          }

\abstract{Intermolecular forces are modeled by means of a modified Lennard-Jones potential, introducing a distance of minimum approach, and the effect of intermolecular interactions is accounted for with a self consistent field of the Vlasov type. A Vlasov equation is then written and used to investigate the propagation of perturbations in a liquid. A dispersion relation is obtained and an effect of damping, analogous to what is known in plasmas as ``Landau damping'', is found to take place.}

\keywords{Vlasov equation \*\ intermolecular potential \*\ modified Lennard-Jones \*\ wave propagation \*\ wave damping}
\pacs{05.60.Cd; 05.20.Dd; 05.20.Jj}

\newcommand{\vect}{\mathbf}
\newcommand{\res}{\mathrm{Re}}
\DeclarePairedDelimiter{\abs}{\lvert}{\rvert}

\begin{document}
\maketitle

\section{Introduction}

The physical behaviour of real gases and liquids is different from that of ideal gases mainly as a consequence of the effect of the intermolecular forces.

Aim of this work is to investigate, in the framework of kinetic theory, the problem of waves propagation in liquids, bringing out the role that intermolecular forces play on the behaviour of wave propagation, in particular on the dispersion relation. Also, there will be shown the existence of an effect of damping, of the type of Landau damping in plasmas, that can be revealed only in the context of microscopic theory, since this effect is strictly related to the form of the distribution function and disappears if oscillations are analyzed in the context of macroscopic theory.
In analyzing waves, one can choose between two methods of approach: either derive the wave equation from macroscopic equations (to wit, the fluid-dynamics equations), or in the framework of kinetic theory, i.e., with a microscopic approach, as is done in the present work.
Here the starting point is the Vlasov equation, particularized for liquids; in this equation the effects of interactions between the molecules of the system is accounted for through a self-consistent field, that in the present case is presented in Appendix, to which the interested reader is referred. The Vlasov equation approach is appropriate for liquids, since molecules therein are subjected to simultaneous interactions with a large number of surrounding molecules, and hence correlation is negligible. This approach is often referred to as mean-field approach (see refs. \cite{r1,r2,r3,r4}; also \cite{r5}, pp. 87 and ff.).

\section{Vlasov equation and wave propagation}

The starting point of the present discussion, as mentioned in the introduction, is the Vlasov equation for the distribution function ${f}(\vect{r},\vect{v},t)$

\begin{equation}
\label{eq:eq1}
 \frac{\partial f}{\partial t}+ \vect v \cdot \frac{\partial f}{\partial \vect r} + \frac{\vect F'}{m} \cdot \frac{\partial f}{\partial \vect v} = 0
\end{equation}

In this equation the global effect of molecular interactions is accounted through the Vlasov self-consistent field $\vect F'$, which is the average effect in the point considered of the forces from all the surrounding particles, weighted on the density distribution of these latter:

\begin{equation}
 \vect F'(\vect{r_1})= \int_{\Re^3} n(\vect{r_2}) \frac{\partial U_{1,2}}{\partial \vect{r_1}} d \vect{r_2} 
\end{equation}

where $ U_{1,2}(\vect{r_1,r_2}) $ is the pairwise interaction potential between two molecules located in $\vect{r_1}$ and  $\vect{r_2}$ respectively and $n(\vect{r_2})$ is the number density at $\vect{r_2}$. Given the above definition, the self-consistent field vanishes in any homogeneous configuration.

\begin{figure}
 \includegraphics[width=0.7\textwidth]{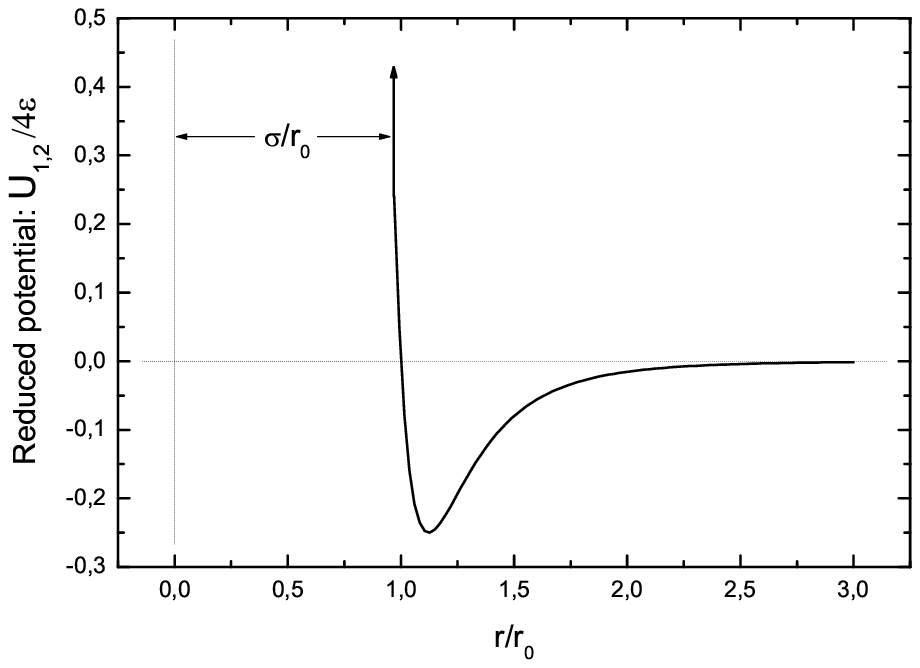}
 \caption{Modified Lennard-Jones intermolecular potential}
 \label{fig:fig1}
\end{figure}

The Vlasov equation is particularly suitable to study wave propagation in a system where the self-consistent field is dominant. In this work, the self-consistent field is derived from a modified Lennard-Jones model (Figure \ref{fig:fig1}),

\begin{equation}
 U_{1,2}(r) =
 \begin{cases}
  \infty & \text{for} \quad \frac{r}{r_0} \leq \frac{\sigma}{r_0} \leq 1 \\
   4 \varepsilon \left[ {\left( \dfrac{r_0}{r} \right)}^{12} - {\left( \dfrac{r_0}{r} \right)}^{6} \right] & \text{for} \quad \frac{r}{r_0}>\frac{\sigma}{r_0}
 \end{cases}
\end{equation}

(where $r=\abs{\vect{r_1}-\vect{r_2}}$), as discussed at large in the Appendix, to which the interested reader is referred for further details. There, the self-consistent field is calculated for the one-dimensional case that will be considered here, as

 \begin{equation}
  F_L(z)\cong \Lambda \frac{d n(z)}{dz}
 \end{equation}

where the parameter $\Lambda$ is given by

\begin{equation}
 \Lambda = \frac{16\pi\varepsilon r_0{^6}}{3\sigma^3} \left[{ 1-{\frac{1}{3}} {\left(\frac{r_0}{\sigma}\right)^6}}  \right]
\end{equation}

and is negative for the present case of a liquid, as discussed in the Appendix.
Consider now a liquid with no external forces (gravity will be neglected here), in some equilibrium described by a distribution function $ f_0(\vect v)$, and a small perturbation defined as follows 

\begin{equation}
 f(z,\vect v,t)=f_0(\vect v)+\varphi(z,\vect v,t)
\end{equation}

and hence

\begin{equation}
 n(z,t)=\int_{\Re^3} f(z,\vect v,t)d_3\nu = \int_{\Re^3} f_0(\vect v)d\vect v + \int_{\Re^3} \varphi(z,\vect v,t)d\vect v =
       n_0 + \eta(z,t) 
\end{equation}

Possible propagation of the perturbation under these conditions will be, if any, along $\hat{\vect{z}}$ the  direction. Introducing the above prescription into (\ref{eq:eq1}) and treating, consistently, the self-consistent field as a perturbation as well, upon neglecting terms of order higher than the first, the Vlasov equation becomes

\begin{equation}
 \frac{\partial \varphi}{\partial t}+\nu_z \frac{\partial \varphi}{\partial z}+\frac{\Lambda}{m}\frac{d\eta}{dz}\frac{\partial f_0}{\partial \nu_z}=0
\end{equation}

Taking a Fourier transform from $z$ to $k$, and a Laplace transform from $t$ to $s$,

\begin{equation}
\label{eq:eq9}
 s\tilde{\tilde{\varphi}}(k,\nu_z,s)+ik\nu_z\tilde{\tilde{\varphi}}(k,\nu_z,s)+ \frac{ik\Lambda N(k,s)}{m} \frac{\partial f_0(\nu_z)}{\partial \nu_z}=\tilde{\varphi_0}(k,\nu_z)
\end{equation}

where $\tilde{\varphi_0}(k,\nu_z)$ is the Fourier transform of $\varphi(z,\nu_z,t)$ at time $t=0$, and $\tilde{\tilde{\varphi}}(k,\nu_z,s)$ and $ N(k,s)$ are the double transforms of the perturbations  $\varphi(z,\nu_z,t)$ and $\eta (z,t)$.

Rearranging (\ref{eq:eq9}) yields

\begin{equation}
 \tilde{\tilde{\varphi}}= \frac{1}{s+ik\nu_z} \left[{\tilde \varphi_0(k,\nu_z)}-{\frac{ik\Lambda N(k,s)}{m}\frac{\partial f_0}{\partial \nu_z}}\right]
\end{equation}

And integration over $\nu_z$ leaves, after rearranging,

\begin{equation}
\label{eq:eq11}
 N(k,s)=\frac{-\frac{i}{k} \int_{-\infty}^{\infty} \frac{\tilde{\varphi_0}(k,\nu_z)}{\nu_z-i\frac{s}{k}}d\nu_z}{1+\frac{\Lambda}{m} \int_{-\infty}^{\infty} \frac{\frac{df_0{\nu_z}}{d\nu_z}}{\nu_z-i\frac{s}{k}}d\nu_z}
\end{equation}

Assuming that the equilibrium distribution function is maxwellian, as is to be expected in the present setting, $f_0(\nu_z)$ is calculated as

\begin{equation}
 f_0(\nu_z)=\int_{\Re{^2}} f_M(\vect v)d\nu_x d\nu_y=n_0{\left(\frac{m}{2\pi K_B T}\right)}^{\frac{1}{2}} exp{\left \{ - \frac{m{\nu_z}^2}{2K_BT} \right \}}
\end{equation}
 
and the derivative inside the integral in the denominator of (\ref{eq:eq11}) becomes

\begin{equation}
\label{eq:eq13}
 \frac{df_0(\nu_z)}{d\nu_z}=n_0 \sqrt{\frac{\beta}{\pi}}2\nu_z\beta e^{-\beta \nu {z^2}}
\end{equation}
 
where $\beta=\frac{m}{2K_{b}T}$.\\ 
Now the integrals in (\ref{eq:eq11}) have to be handled in the complex plane, and the complex extension of the variable $\nu_z$ will be noted in the following as $w$. The above integrals will then be written as path integrals, and (\ref{eq:eq11}) rewritten as follows

\begin{equation}
\label{eq:eq14}
  N(k,s)=\frac{-\frac{i}{k} \int_{\Gamma} \tilde{\varphi_0}(k,w) \frac{dw}{w - i\frac{s}{k}}}{1+\frac{\Lambda}{m} \int_{\Gamma} \frac{df_0(w)}{dw}\frac{dw}{w - i\frac{s}{k}}}
\end{equation}

where the path $\Gamma$ is the straight line that lies on the real axis. The function $\frac{df_0(w)}{dw}$, complex extension of (\ref{eq:eq13}), is analytic everywhere, and $\tilde{\varphi_0}(k,w)$ can be assumed to be well behaved as well, for physical systems: if this is the case, the only singularity in either integrand is the simple pole in $w=\frac{is}{k}$. Without attempting Laplace inverse transformation of (\ref{eq:eq14}), it is noted that the result in (\ref{eq:eq14}) parallels that found by Landau \cite{r6}, allowance made for the factor in front of the integral in the denominator and for the specific form of the equilibrium distribution function $f_o(w)$: the same procedure can be followed, leading to the asymptotic solution \cite{r5,r6,r7}. At large enough times, the evolution is dominated by the rightmost pole of (\ref{eq:eq14}), located in the rightmost zero of its denominator, call it $s_1$, producing asymptotically a simple exponential behaviour:

\begin{equation}
 \tilde{\eta}(k,t) \propto exp {\left \{ s_1t \right \}}
\end{equation}
 
\begin{figure}
 \includegraphics[width=0.7\textwidth]{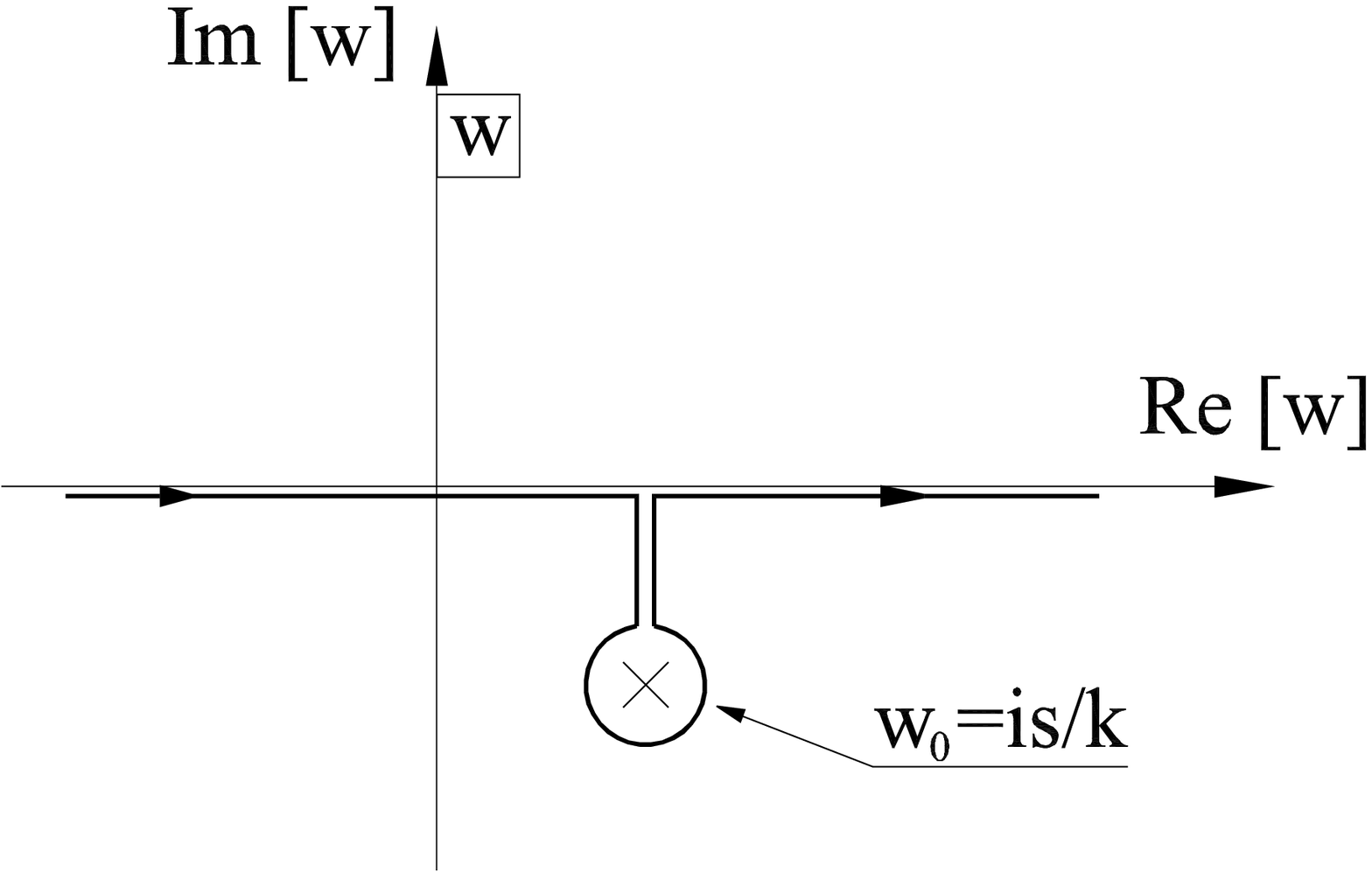}
 \caption{Integration contour for $\res[i\frac{s}{k}]\leqslant 0$}
 \label{fig:fig2}
\end{figure}

If the real part of $s_1$ is negative, it produces a ``Landau-type'' damping of the propagation. Now, $s_1$ is a solution of the following equation

\begin{equation}
\label{eq:eq16}
 D(k,s)= 1 + \frac{\Lambda}{m} \int_{\Gamma} \frac{df_0(w)}{dw} \frac{dw}{w - i\frac{s}{k}}= 1-\frac{\Lambda2n_0 \beta ^{\frac{3}{2}}}{m \sqrt{\pi}} \int_{-\infty}^{\infty} \frac{we^{-\beta w^2}}{w - i\frac{s}{k}} dw =0
\end{equation}
 
where the result of integration is a complex function of the complex variable s. Equation (\ref{eq:eq16}) is the dispersion relation of the propagation. 
To investigate the asymptotic behavior, it is necessary to define an analytic extension to the entire complex plane s of the functions resulting from the path integral: this is effected through the choice of the contour shown in Figure \ref{fig:fig2} for the integration path $\Gamma$.

When $\res[s]<0$ , and the pole in $w_o=i\frac{s}{k}$ takes on values located on or below the real axis, the integration path is be deformed to circle the singularity, and calculation of the integral yields two terms: the Cauchy principal value and the residue in the pole. Hence, it is convenient to look for the solution to (\ref{eq:eq16}) separately in the right half plane $\res[s]>0$ and in the left one $\res[s]<0$.
The analysis will be conducted with reference to oscillations with wavelength such that the wave phase velocity is larger than the mean thermal velocity, that is

\begin{equation}
\label{eq:eq17}
 \frac{Im[s]}{k}>>\nu_{th}=\sqrt{\frac{2K_BT}{m}}=\beta^{-\frac{1}{2}}
\end{equation}

This hypothesis is introduced on account of the physical consideration that if the mean thermal velocity is larger than the phase velocity, the random motion prevails on the orderly motion of the wave and the wave does not propagate. 

\subsection{Right half plane: $\res[s]>0$}

Calculating the principal value of the integral, (\ref{eq:eq16}) becomes 

\begin{equation}
\label{eq:eq18}
 1-\frac{2\Lambda n_0\beta}{m}+\frac{2\Lambda n_0\sqrt{\pi}\beta^{\frac{3}{2}}s}{mk}erfc(\sqrt{\beta}\frac{s}{k})e^{\beta \frac{s^2}{k^2}}=0
\end{equation}

where $erfc$ is the complementary error function. In the hypothesis of (\ref{eq:eq17}), the $erfc$ function can be conveniently expanded as follows \cite{r8}

\begin{equation}
 \sqrt{\pi}ze^{z^2}erfc(z)=1-\frac{1}{2z^2}+\frac{3}{4z^4}+O {\left[ \frac{1}{z^6} \right]}
\end{equation}
 
and inserting this expansion into (\ref{eq:eq18}) yields

\begin{equation}
\label{eq:eq20}
 1-\frac{\Lambda n_0k^2}{ms^2}+\frac{3\Lambda n_0k^4}{2m\beta s^4}=0
\end{equation}
 
By simple inspection, $\frac{s}{k}=0$ is not a root of (\ref{eq:eq20}), so the equation can be multiplied throughout by $(\frac{s}{k})^4$  yielding finally

\begin{equation}
\label{eq:eq21}
 {\left( \frac{s}{k} \right)}^4-\frac{\Lambda n_0}{m} {\left(\frac{s}{k} \right)}^2 +\frac{3\Lambda n_0}{2m\beta}=0
\end{equation}

The complex variable $s$ may be written as $s=\omega(\alpha+i)$: then, in the hypothesis $|\alpha|<<1$ 

\begin{equation}
\label{eq:eq22}
 s^2=\omega^2(\alpha+i)^2\approx(-1+i2\alpha)\omega^2
\end{equation}
 
\begin{equation}
\label{eq:eq23}
 s^4=\omega^4(\alpha+i)^4\approx(1-i4\alpha)\omega^4
\end{equation}

Introducing these values into (\ref{eq:eq21})

\begin{equation}
\label{eq:eq24}
 {\left(\frac{\omega}{k}\right)}^4(1-i4\alpha)-{\left(\frac{\omega}{k}\right)}^2\frac{\Lambda n_0}{m}(1-i2\alpha)+\frac{3\Lambda n_0}{2m\beta}=0
\end{equation}
 
Separating real and imaginary parts, the following two equations are obtained

\begin{equation}
\label{eq:eq25}
 {\left(\frac{\omega}{k}\right)}^4-{\left(\frac{\omega}{k}\right)}^2\frac{\Lambda n_0}{m}+\frac{3\Lambda n_0}{2m\beta}=0
\end{equation}

\begin{equation}
\label{eq:eq26}
 -4\alpha{\left(\frac{\omega}{k}\right)}^4+2\alpha{\left(\frac{\omega}{k}\right)}^2\frac{\Lambda n_0}{m}=0
\end{equation}
 
From (\ref{eq:eq25}) the phase velocity can be calculated:

\begin{equation}
 {\left(\frac{\omega}{k}\right)}^2=-\frac{\Lambda n_0}{2m}\left[1+{\sqrt{1-\frac{6m}{\beta\Lambda n_0}}}\right]=0
\end{equation}
 
where, again, it should be born in mind that the parameter $\Lambda$ is negative, as discussed in the Appendix. Introducing it into (\ref{eq:eq26}) 

\begin{equation}
 \alpha\left[{\frac{\Lambda n_0}{m}}-{2\left(\frac{\omega}{k}\right)^2}\right]=\alpha\frac{\Lambda n_0}{m}\left[1+\left[1+{\sqrt{1-\frac{6m}{\beta\Lambda n_0}}}\right]\right]=0
\end{equation}

from which $\alpha=0$, in other words, there is no solution with $\res[s]>0$.

\subsection{Left half plane: $\res[s]<0$}

In this case from (\ref{eq:eq21}) and (\ref{eq:eq24}) and following the same procedure as in the case $\res[s]>0$ 

\begin{equation}
 D(s,k)=1-\frac{\Lambda n_0k^2}{ms^2}+\frac{3\Lambda n_0k^4}{2m\beta s^4}+\frac{4\Lambda n_0\beta^{\frac{3}{2}}\sqrt{\pi}s}{mk}e^{\frac{\beta s^2}{k^2}}=0
\end{equation}

Introducing the approximation of (\ref{eq:eq22}) and (\ref{eq:eq23}) and recalling the hypothesis $|\alpha|<<1$ so that 

\begin{equation}
 e^{\frac{\beta s^2}{k^2}} \approx e^{-\beta{\frac{\omega^2}{k^2}}} e^{i2\beta \alpha{\frac{\omega^2}{k^2}}}
\end{equation}

the real part of the dispersion equation becomes

\begin{equation}
 {\left(\frac{\omega}{k}\right)}^4+{\left(\frac{\omega}{k}\right)}^2\frac{\Lambda n_0}{m}+\frac{3\Lambda n_0}{2m\beta}+{\left(\frac{\omega}{k}\right)}^5\frac{\Lambda n_04\sqrt{\pi}\beta^{\frac{3}{2}}}{m}e^{-\beta\left(\frac{\omega}{k}\right)^2}\left \{ {5\alpha \cos \left(2\beta \left(\frac{\omega}{k} \right)^2 \alpha \right)}-{ \sin \left(2\beta \left(\frac{\omega}{k} \right)^2 \alpha \right)} \right \}
\end{equation}

Now, observing that the expression into parenthesis $\approx 0$ , the relation between $\omega$ and $k$ that gives the waves that can propagate in liquids, is the same as (\ref{eq:eq25}).
For the imaginary part, one obtains 

\begin{equation}
 \alpha\frac{2\Lambda n_0k^2}{m} \left({\frac{3k^2}{\beta}}+{\omega^2}\right)+\frac{4\Lambda _0\beta^{\frac{3}{2}}\omega^5}{mk}e^{-\frac{\beta\omega^2}{k^2}}\left \{ { \cos \left( {\frac{\beta2\omega^2\alpha}{k^2}} \right)}+{ \alpha \sin \left({\frac{\beta2\omega^2\alpha}{k^2}}\right)} \right \}=0
\end{equation}

The expression into parenthesis is $\approx 1$ and then the value of $\alpha$ is given by

\begin{equation}
 \alpha \approx -\frac{\beta^{\frac{5}{2}}\omega^5e^{-\frac{\beta\omega^2}{k^2}}}{\left(3+{\frac{\beta\omega^2}{k^2}} \right)}
\end{equation}

\section{Conclusion}

In the present work, results are obtained from a kinetic theory approach, in particular: the dispersion relation; the phase velocity as a function of the intermolecular force; the ``Landau-type'' damping effect, i.e., a damping analogous to that encountered in plasmas, which can also only be seen from a kinetic approach. Starting from kinetic theory, some aspects are seen that cancel out when a macroscopic equations approach is followed.

\appendix*
\section{}\label{appA}

In the Vlasov equation \cite{r7,r9}

\begin{equation}
 \frac{\partial f}{\partial t}+\vect v\cdot\frac{\partial f}{\partial\vect r}+\frac{\vect F'}{m}\cdot\frac{\partial f}{\partial \vect v}=0
\end{equation}

the effect of molecule interaction is accounted for through a self-consistent field $\vect F'$, to be calculated as

\begin{equation}
\label{eq:eq_a2}
 \vect F'(\vect r)=\int_V n(\vect r_2)\vect F_{1,2}d\vect r_2
\end{equation}

where $\vect F_{1,2}$ is the force that a molecule located at position $\vect r_2$ exerts on the molecule in $\vect r$ and $n(\vect r)$ is the local number density at $\vect r$.
The detailed form of the interaction function can be investigated only through quantum mechanics and much work has been done in this direction \cite{r10,r11}. However the problem is very complex and many effects are involved; moreover the structure of the molecules is often not very well known. Therefore the existing results contain significant approximations and are applicable only to specific situations. This being the case, it becomes essential to resort to a phenomenological potential $U_{1,2}$.
In this work, the following Lennard-Jones model, modified to include a distance of closest approach $\sigma$ to account for the non-vanishing dimensions of the molecules, will be used to calculate the self-consistent field. The intermolecular potential $U_{1,2}(r)$ (henceforth referred to as modified Lennard-Jones model or mLJ) is presented in Figure \ref{fig:fig1}, and is given by

\begin{equation}
 U_{1,2}(r) =
 \begin{cases}
  \infty & \text{for} \quad r \leq \sigma \leq r_0 \\
   4 \varepsilon \left[ {\left( \dfrac{r_0}{r} \right)}^{12} - {\left( \dfrac{r_0}{r} \right)}^{6} \right] & \text{for} \quad r>\sigma
 \end{cases}
\end{equation}

\begin{figure}
 \includegraphics[width=0.6\textwidth]{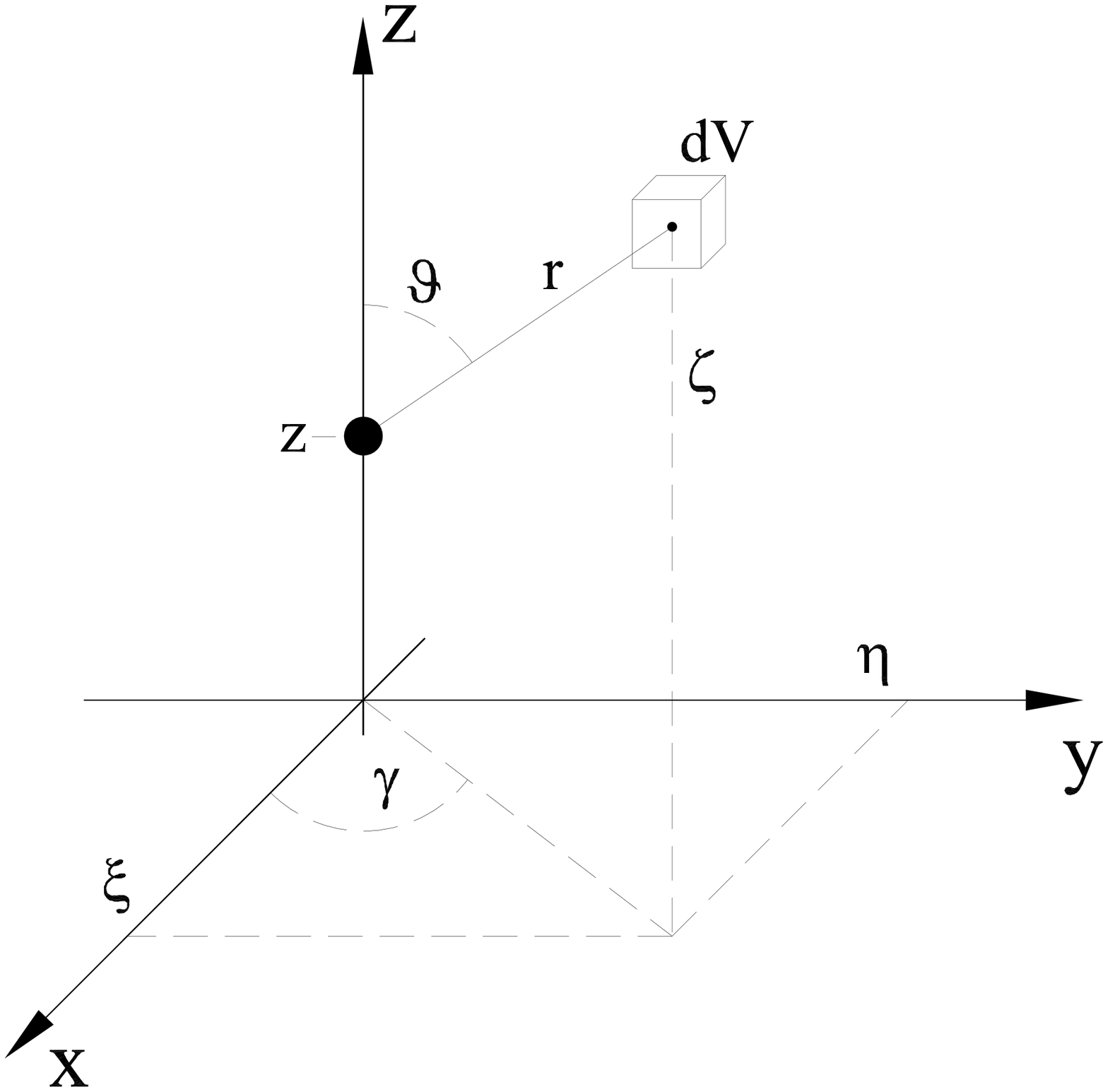}
 \caption{Calculation geometry}
 \label{fig:fig_a1}
\end{figure}

To calculate the self-consistent force $\vect F'$ , a molecule located at the point $(0,0,z)$ will be considered, and the force exerted on this by the whole surrounding liquid will be calculated from the mLJ potential.
To simplify the problem, a system possessing slab symmetry will be assumed, that is one in which density depends only on the z - coordinate. Consider then an elementary volume $dV$ at a location defined by the coordinates $(r,\vartheta,\gamma)$ in a spherical reference system centered in the molecule of interest and with the polar axis along the $z$ direction, see Figure \ref{fig:fig_a1} \cite{r12}

With the geometry in Figure \ref{fig:fig_a1} the force acting on the molecule of interest due to a molecule in $(r,\vartheta,\gamma)$ becomes

\begin{equation}
 \vect F_{1,2}=4\varepsilon \left[\frac{6r_0^6}{r^7}-\frac{12r_0^{12}}{r^{13}}\right] \vect{\hat{r}} \quad for \quad r>\sigma 
\end{equation}
 
Now, calling $(\xi,\eta,\zeta)$ the cartesian coordinates of volume $dV$, the value of $r^{\alpha+1}$ with $\alpha=12$ or $\alpha=6$ can be calculated as 

\begin{equation}
 r^{\alpha+1}=\left[{\xi^2}+{\eta^2}+{(\zeta-z)^2}\right]^{\frac{\alpha+1}{2}}
\end{equation}

Sines and cosines of the angles in Figure \ref{fig:fig_a1} can be expressed in terms of the cartesian coordinates $(\xi,\eta,\zeta)$

\begin{equation}
\label{eq:eq_a3}
 \begin{cases}
  \sin\vartheta=\frac{\sqrt{\xi^2+\eta^2}}{\sqrt{\xi^2+\eta^2+(\zeta-z)^2}} & \sin\gamma=\frac{\eta}{\sqrt{\xi^2+\eta^2}} \\
  \cos\vartheta=\frac{\zeta-z}{\sqrt{\xi^2+\eta^2+(\zeta-z)^2}} & \cos\gamma=\frac{\xi}{\sqrt{\xi^2+\eta^2}} \\
 \end{cases}
\end{equation}

The Cartesian component of the force may be rewritten as

\begin{equation}
 F_{1x}=4\varepsilon n(\zeta)d\xi d\eta d\zeta \left[\frac{\xi r_0^{\alpha}\alpha}{\left[\xi^2+\eta^2+(\zeta-z)^2\right]^{\frac{\alpha+2}{2}}} \right]_{\alpha=12}^{\alpha=6}
\end{equation}

\begin{equation}
 F_{1y}=4\varepsilon n(\zeta)d\xi d\eta d\zeta \left[\frac{\eta r_0^{\alpha}\alpha}{\left[\xi^2+\eta^2+(\zeta-z)^2\right]^{\frac{\alpha+2}{2}}} \right]_{\alpha=12}^{\alpha=6}
\end{equation}

\begin{equation}
 F_{1z}=4\varepsilon n(\zeta)d\xi d\eta d\zeta \left[\frac{(\zeta-z)r_0^{\alpha}\alpha}{\left[\xi^2+\eta^2+(\zeta-z)^2\right]^{\frac{\alpha+2}{2}}} \right]_{\alpha=12}^{\alpha=6}
\end{equation}

To obtain the overall force on the reference molecule, integration over the whole volume is performed. It can be seen readily that 

\begin{equation}
 \int_{-\infty}^{\infty}\int_{-\infty}^{\infty} F_{1x}d\xi d\eta=\int_{-\infty}^{\infty}\int_{-\infty}^{\infty} F_{1y}d\xi d\eta=0
\end{equation}

so that there are no x and y components to the force – and this is consistent with the symmetry of the problem. As for the z component

\begin{equation}
 \int_{-\infty}^{\infty}\int_{-\infty}^{\infty} F_{1z}d\xi d\eta=n(\zeta)d\zeta(\zeta-z) \left[\int_{-\infty}^{\infty}\int_{-\infty}^{\infty} \frac{4\varepsilon\alpha r_0^\alpha}{\left[\xi^2+\eta^2+(\zeta-z)^2\right]^{\frac{\alpha+2}{2}}} d\xi d\eta \right]_{\alpha=12}^{\alpha=6}
\end{equation}

Considering the mLJ potential (\ref{eq:eq_a3}), there is a minimum approach distance $\sigma$

\begin{equation}
\label{eq:eq_a12}
 F_L=8\pi\varepsilon \left[{-r_0^\alpha \int_{-\infty}^{z-\sigma} \frac{n(\zeta)}{(z-\zeta)^5}d\zeta}+{r_0^\alpha \int_{z+\sigma}^{\infty} \frac{n(\zeta)}{(z-\zeta)^5}d\zeta} \right]_{\alpha=12}^{\alpha=6}
\end{equation}

If the density variation is mild, $n(z)$ can be expanded in Taylor series retaining only the first few terms 

\begin{equation}
 n(\zeta)=n(z)+\frac{dn(z)}{dz}(\zeta-z)+\frac{d^2n(z)}{dz^2}\frac{(\zeta-z)^2}{2}+\frac{d^3n(z)}{dz^3}\frac{(\zeta-z)^3}{3!}+\frac{d^4n(z)}{dz^4}\frac{(\zeta-z)^4}{4!}+O\left[(\zeta-z)^5\right]
\end{equation}

Neglecting terms of order 5 and higher, and substituting into (\ref{eq:eq_a12}), after some algebra the following equation is obtained:

\begin{equation}
\label{eq:eq_a14}
 F_L(z)\cong \Lambda\frac{dn(z)}{dz}+\Lambda_3\frac{d^3n(z)}{dz^3}
\end{equation}

where the coefficients are given by

\begin{equation}
 \Lambda=\frac{16\pi\varepsilon r_0^6}{3\sigma^3}\left[1-{\frac{1}{3}\left(\frac{r_0}{\sigma} \right)^6} \right] \qquad \Lambda_3=\frac{8\pi\varepsilon r_0^6}{3\sigma}\left[1-{\frac{1}{7}\left(\frac{r_0}{\sigma} \right)^6} \right]
\end{equation}

In the present work, only the first term in (\ref{eq:eq_a14}), i.e., $\Lambda$, will be retained, yielding

\begin{equation}
 F_L(z)\cong \frac{16\pi\varepsilon r_0^6}{3\sigma^3}\left[1-{\frac{1}{3}\left(\frac{r_0}{\sigma} \right)^6}\frac{dn(z)}{dz} \right] 
\end{equation}

The question arises of the sign of $\Lambda$, which depends on the value of the $\frac{r_0}{\sigma}$ ratio: if this is larger than $\sqrt[6]{3}$ (i.e., approximately 1.20), $\Lambda$ becomes negative. Now the value of the distance of closest approach $\sigma$ is dependent on the temperature and the density \cite{r13}, however phenomenological considerations may yield a qualitative answer: liquids do not yield easily to compression, and on the other hand have a strong cohesive behaviour; departures from their equilibrium density at the given temperature is strongly resisted. This leads one to believe that the coefficient $\Lambda$ needs to be negative, so that the force opposes the density gradient, trending to restore the equilibrium density.

\end{document}